\magnification=1200
\parskip=10pt plus 5pt
\parindent=14pt
\baselineskip=12pt
\input amssym.def
\input amssym
\topglue 0.6in
\pageno=0
\footline={\ifnum \pageno < 1 \else \hss \folio \hss \fi}
\line{\hfil{Report 99-12~(Revised)~~~~}}
\line{\hfil{November, 1999~~~~~~~~~~~~~}}
\topglue 1in
\centerline{\bf HIGHER-DERIVATIVE QUANTUM COSMOLOGY}
\vskip .5in
\centerline{\bf Simon Davis}
\vskip 20pt
\centerline{School of Mathematics and Statistics}
\centerline{University of Sydney}
\centerline{NSW 2006}
\vskip 30pt
\centerline{\bf Abstract}  
The quantum cosmology of a higher-derivative gravity theory arising
from the heterotic string effective action is reviewed.  A new type of 
Wheeler-DeWitt equation is obtained when the dilaton is coupled 
to the quadratic curvature terms.  Techniques for solving 
the Wheeler-DeWitt equation with appropriate boundary conditions shall
be described, and implications for semiclassical theories of inflationary 
cosmology will be outlined.
\vskip 30pt
\noindent
{\bf Current version of the text of a talk given at ACGRG2, 9 July 1998}

\vfill\eject

\noindent
{\bf 1. Higher-Derivative Gravity Theories}

Since the renormalization of the Einstein-Hilbert action for general 
relativity introduces higher-order curvature terms, it has long been
expected that the development of quantum cosmology would 
include higher-derivative gravity theories.  Further support for this
approach arises from the presence of quadratic and higher-order curvature
terms in string effective actions, where modifications to the
Ricci scalar are required for maintaining conformal invariance of the sigma 
model, describing string propogation in a given background,
even after inclusion of quantum corrections to the $\beta$-function.

Typically, one may consider actions of the form
$$\eqalign{I~&=~\int~d^4x~{\sqrt {-g}}~(R+c_1(g_{\mu\nu},\phi,\chi_\gamma,...)
(R_{\mu\nu\rho\sigma}R^{\mu\nu\rho\sigma}
+\alpha R_{\mu\nu}R^{\mu\nu}~+~\beta~R^2)+O(R^3))
\cr}
\eqno(1)
$$
and field-dependence of the coefficients $c_1,...$ provides couplings between
matter fields and the curvature.

There are several types of higher-derivative gravity theories selected
by physical considerations:
\vskip 5pt
\noindent
(i) Lovelock gravity required for unitarity and elimination of ghosts
\vskip 2pt
Amongst the curvature invariants that arise in superstring effective actions
are the
\hfil\break
dimensionally continued Lovelock invariants
$$L_{(n)}~=~{{(2n)!}\over {2^n}}~R_{[i_1i_2}{}^{i_1i_2}R_{i_3i_4}{}^{i_3i_4}
...R_{i_{2n-1}i_{2n}]}{}^{i_{2n-1}i_{2n}}
\eqno(2)
$$
which vanish in ten dimensions when $n\ge 6$. As generalizations
of the Gauss-Bonnet invariant, they are the only combinations of Riemann 
tensors that give rise to second-order field equations for the metric.  In a
sigma-model with expansion parameter $\alpha^\prime$, these curvature
terms would occur at order $O(\alpha^{n-1})$ in the effective action.
The elimination of ghost fields in higher-order gravitational actions is 
achieved by eliminating terms of the type $h\nabla^{2n} h,~n~\ge~2$, 
where $h_{\mu\nu}$ is the perturbation about the flat-space metric, through
metric redefinitions.  The linearized equation for the graviton field
is then second-order, possessing physically
consistent stable classical solutions, in contrast to equations with third and
higher-order derivatives of the metric, and at the quantum level, the
ghost poles will cancel in the graviton propagator.
The only non-zero Lovelock terms that could appear in ten-dimensional
superstring effective actions are $L_{(0)},~L_{(1)},~L_{(2)},~L_{(3)},
~L_{(4)},~L_{(5)}$ and functions of these tensors. While $L_{(5)}$ is a
topological invariant in ten dimensions, the other three curvature
combination $L_{(2)},~L_{(3)},~L_{(4)}$ contain dynamical higher-derivative 
terms.
\vfill\eject
\noindent
(ii) Superstring effective actions with higher-order curvature terms
\vskip 2pt
There is a particular combination of Lovelock invariants which is 
equivalent to a Chern-Simons theory with SO(1,9) gauge group [1], 
and this suggests the existence of a ten-dimensional superstring theory 
reducing to an effective field theory containing D=10 super-Yang-Mills 
theory with gauge group SO(1,9), combined with topological terms, as an 
appropriate modification of general relativity with improved renormalizability
properties.

In this superstring effective action, the curvature terms are 
generated by identifying the gauge potential with the metric connection. 
The various different quartic curvature combinations that can be included
in low-energy superstring  and heterotic effective string actions are
restricted by their consistency with supersymmetry [2][3].  

\vskip 5pt
\noindent
(iii) Born-Infeld actions
\vskip 2pt
Born-Infeld effective actions, which typically involve determinants of 
two-index fields such as the field strength $F_{\mu\nu}$ or the 
Riemann tensor $R_{\mu\nu}$ may represent a viable alternative to the 
expansion of the effective action in powers of $\alpha^\prime$, relevant for
specific physical configurations.  Difficulties associated with the
infinite expansion in powers of $\alpha^\prime$ at the Planck scale
might then be circumvented.
\vskip 10pt
\noindent{\bf 2. Classical Cosmological Solutions of Equations of
Motion}
\hfil\break
\phantom{.....}{\bf of Higher-Derivative Theories}

Since solutions to the general relativistic field equations  
contain initial curvature singularities whenever the dominant
energy condition is satisfied, one of the motivations for developing
quantum cosmology has been the theoretical justification of the 
absence of the singularity.  Non-singular solutions may arise in
theories which are modifications of general relativity, and          
classical cosmological solutions to the equations of motion for several 
different types of theories containing higher-order curvature 
terms have been analyzed with regard to the absence 
of singularities and the existence of de Sitter phases.

\vskip 5pt
\noindent
(i) Singularity-free cosmological solutions of heterotic string
effective actions with quadratic 
\phantom{....}curvature terms
\vskip 2pt
An action, which combines higher-derivative gravity with a 
scalar field, has been shown to have singularity-free 
cosmological solutions [4][5].
At string tree-level and first-order in the $\alpha^\prime$-expansion
of the compactified heterotic string effective action in four dimensions,
the dynamics of the graviton, dilaton field S and modulus field T can be
described by
$$L_{eff.}~=~{1\over {2\kappa^2}}R~+~{{DS D{\bar S}}\over {(S+{\bar S})^2}}
~+~3{{DT D{\bar T}}\over {(T+{\bar T})^2}}
~+~{1\over 8}~(Re~S)~R_{GB}^2~+~{1\over 8}(Im~S)~R {\tilde R}
\eqno(3)
$$
The couplings for the quadratic curvature terms are dimensionless,
and this is consistent with the $\alpha^\prime$ expansion because
${{\alpha^\prime}\over {\kappa^2}}$ has no units in four dimensions.
If $Re~T$, representing the square of the compactification radius, is
set equal to a constant, and $Im~T~=~0$, the kinetic term 
for the modulus field vanishes.  In addition, defining the real part of the
dilaton field to be $Re~S~=~{1\over {g_4^2}}~e^\Phi$, 
setting $Im~S$ equal to zero and choosing units such that 
$\kappa~=~1$ and the effective action can be set equal to
$$I~=~\int~d^4x~{\sqrt{-g}}~\left[~R~+~{1\over 2}(D\Phi)^2~+~
{{e^\Phi}\over {4g_4^2}}(R_{\mu\nu\kappa\lambda} R^{\mu\nu\kappa\lambda}
~-~4~R_{\mu\nu} R^{\mu\nu}~+~R^2)\right]
\eqno(4)
$$
This action is also obtained when $Im~S~=constant$ and the background 
geometry is restricted to have a Friedmann-Robertson-Walker metric.
The Gauss-Bonnet invariant arises in this action,
but it is multiplied now by the factor 
${{e^\Phi}\over {4 g_4^2}}$, where $g_4$ is the four dimensional
string coupling constant and $\Phi$ is a scalar field, so that the
integral is not a topological invariant.

In the phase space of classical solutions to the field equations of the
four-dimensional action (4), there is a class of space-times which
have no singularity for a large range of values of the dilaton and 
modulus fields.  
\vskip 2pt
\noindent
(ii) Dimensionally-continued Euler actions and Lovelock gravity
theories
\vskip 5pt
The study of cosmological solutions of Kaluza-Klein theories [6] and string 
effective actions has led to investigations of dimensionally-continued
Euler actions and Lovelock theories [7][8].  The direct product of a  
four-dimensional Friedmann-Robertson-Walker metric and a metric for a 
compact six-dimensional space with a second scale factor is postulated.  
The gravitational field equations may be solved to obtain the 
time-dependence of the scale factors.  The geodesic completeness of
cosmological solutions to Kaluza-Klein theories with quadratic curvature 
terms has also been investigated [9].  
\vskip 2pt
\noindent
(iii) $R^2$ inflation and higher-derivative gravity without scalar fields
\vskip 5pt
An early analysis in the previous decade of $R^2$ theories [10] and $C^2$
theories [11] showed that an $R^2$ term leads to particle production and 
inflation with minimal dependence on the initial conditions.  The $C^2$
term gives rise to large anisotropy [12] and the destabilization of
positive $\Lambda$ metrics.  Inflation has also been derived from 
higher-derivative terms directly obtained as renormalization 
counterterms [13][14] without the inclusion of scalar or inflaton fields. 
\vfill\eject
\noindent
(iv) Higher-derivative gravity with the dilaton field
\vskip 2pt
Dilaton fields have been considered useful for the cosmology to 
exhibit inflation.  It is known that most classical string
equations of motion do not lead to inflation [15].  A set of 
higher-derivative gravity theories with a dilaton field has been 
investigated and shown to produce the required inflationary growth of the 
Friedmann-Robertson-Walker scale factor [16].  

\noindent
(v) Cosmology of M-type theories
\vskip 2pt
More recently, the low-energy effective actions derived with
the higher-dimensional M-theories have been studied [17][18].  The actions
generally involve the coupling of standard gravity to scalar fields
with a potential determined by the theory, although the analysis could be
extended to include higher-orders curvature terms.  Classical
cosmological solutions have been obtained, by compactifying on flat or
maximally symmetric subspaces [17], and their singularity structure
has been determined.
\vskip 10pt
\noindent{\bf 3. Quantum Cosmology for Gravity Plus Higher-Order Terms}
\vskip 5pt
Much of the initial work on higher-derivative quantum cosmology has been
developed with only curvature terms and no scalar field in the action.
The quantum cosmology of standard gravity coupled to a scalar field has been 
investigated by many authors [19]-[22].  These techniques have
been adapted to string cosmology, based on an effective action consisting of 
the Ricci scalar, dilaton field and antisymmetric 3-index 
field [23]-[27]. The non-zero vacuum
expectation value of the scalar field in the potentials 
arising in grand unified theories drives inflation in semi-classical
cosmology and again it is found to be useful in obtaining wave
functions representing inflationary solutions in quantum cosmology.   
\vskip 5pt
\noindent
(i) Without scalar fields
\vskip 2pt
The quantum cosmology of superstring and heterotic string effective actions in 
ten dimensions with higher-derivative curvature terms up to fourth order also 
has been investigated [28][29].  The Wheeler-DeWitt equations for both 
theories, in the mini-superspace of metrics with different scale 
factors for the physical
and internal spaces, differ significantly from the equation
obtained for the higher-derivative theory considered in this paper
because of the coupling of the curvature and scalar field.  For the
superstring, the differential equation is not generally solvable by
analytic methods, and it is only reducible to the form of a diffusion
equation when the curvatures of the physical and internal spaces are
set equal to zero, and the scalar field is set equal to a constant.
\vfill\eject
\noindent
(ii) With scalar fields
\vskip 2pt
The coupling of scalar fields to higher-derivative curvature terms in 
string effective actions results in Hamiltonians which are not quadratic
function of the momenta.
Even when there are only quadratic curvature terms in the heterotic string 
effective action, the Hamiltonian cannot be expressed 
as a simple function of the canonical momenta, preventing a 
derivation of the Wheeler-DeWitt equation [29].  The derivatives 
of the coordinate fields for the model studied later in
this paper are also found to be given by expressions containing
fractional roots and inverses of canonical momenta.  However, it is shown 
that resulting equation can be converted to a partial differential equation,
resolving the problem of deriving a Wheeler-DeWitt equation.
 
Higher-order terms in the effective action will give rise to corrections in 
the theoretical predictions for the inflationary epoch.  
Given a fundamental theory at Planck scale 
with higher-order terms, it is 
appropriate to consider a boundary located between the Planck era and
the inflationary epoch where the predictions of quantum cosmology of the
higher-derivative theory could be matched, in principle, to the predictions
of the quantum theory of standard gravity coupled to matter fields.
The inclusion of this boundary will have an effect on both the
quantum cosmology of the more fundamental theory and the computations
of the standard model.
\vskip 5pt
\noindent
(iii) $f(R)$ theories
\vskip 2pt
The quantum cosmology of $f(R)$ theories can be contrasted with that of
quadratic gravity theories [30].  The essential simplification in the study
of these theories is a conformal transformation which maps the $f(R)$ theory
to an Einstein-Hilbert action coupled to a scalar field.  A cubic curvature
term, for example, has been found to lead to the existence of a region in
parameter space for which neither the no-boundary or tunneling boundary
conditions produce an inflationary growth that simultaneously resolves 
the horizon and flatness problems.
\vskip 5pt
\noindent
{\bf 4. Supersymmetric Quantum Cosmology with Higher-Derivative Terms}
\vskip 2pt
The model (4) containing quadratic curvature terms and the dilaton field
can be quantized and the Hamiltonian constraint would be given by
the Wheeler-DeWitt equation.  Since the solution to this equation generally
requires a reduction in the number of degrees of freedom in the metric
field, it is convenient to consider only a minisuperspace
of Friedmann-Robertson-Walker metrics with $K=1$ (closed model), 
$K=0$ (spatially flat model) or $K=-1$ (open model). 
The minisuperspace action is then
$$\eqalign{I~&=~\int~d^4 x~a^3(t)~{{r^2~sin~\theta}\over {(1-Kr^2)^{1\over 2}}}
~\biggl[6a^{-2}(a{\ddot{a}}~+~{\dot a}^2~+~K)~+~{1\over 2}~(D \Phi)^2
\cr
&~~~~~~~~~+~{{e^\Phi}\over {4 g_4^2}}~\biggl(-18 {{\ddot a}\over a}^2  
- {6\over {a^4}} (a{\ddot a}
~+~2{\dot a}^2~+~2 K)^2~+~24~a^{-2}~\left({\ddot a}~+~{{{\dot a}^2}\over a}
+~{K\over a}\right)^2~\biggr)\biggr]
\cr}
\eqno(5)
$$
When a boundary is placed for $K~=~1$, and the action 
reduces to a one-dimensional integral
$$\eqalign{I~=~{\bar V}~\int~dt~\biggl[(6a^2 {\ddot a}~&+~6a{\dot a}^2~+~6aK)~
+~{1\over 2}
a^3(D\Phi)^2~+~6{{e^\Phi}\over {4 g_4^2}}{\ddot a}
({\dot a}^2~+~K)\biggr]
\cr}
\eqno(6)
$$
where ${\bar V}$ is a time-independent volume factor, which is given by
${{V(t_f)}\over {a^3(t_f)}}$, where $V(t_f)$ is the volume of the 
three-dimensional spatial hypersurface at a fixed final time $t_f$.

The one-dimensional integral (6) is an example of an action of the following
type
$$I~=~\int~dt~L(t,y,{\dot y},...,y^{(m)}, z, {\dot z}, ..., z^{(n)})
\eqno(7)
$$
The conjugate momenta are defined to be
$$\eqalign{p_1~&=~{{\partial L}\over {\partial {\dot y}}}~-~ 
{d\over {dt}}\left(
{{\partial L}\over {\partial {\ddot y}}}\right)~+~...~+~(-1)^{m-1}
{{d^{m-1}}\over {dt^{m-1}} }\left({{\partial L}\over {\partial y^{(m)}}}
\right)
\cr
p_2~&=~~~~~{{\partial L}\over {\partial {\ddot y}}}~-~{d\over {dt}}
\left({{\partial L}\over {\partial y^{(3)}}}\right)~+~...
~+~(-1)^{m-2}~{{d^{m-2}}\over {dt^{m-2}}}\left({{\partial L}\over {\partial
y^{(m)}}}\right)
\cr
&~\vdots
\cr
p_m~&=~~~~~~~~~~~~~~~~~~~~~~~~~~~~~~~~~~~~~~~~~~~~~~~~~~~~~~~~~~~~
{{\partial L}\over {\partial y^{(m)}}}
\cr
p_{m+1}~&=~{{\partial L}\over {\partial {\dot z}}}~-~{d\over {dt}}
\left({{\partial L}\over {\partial {\ddot z}}}\right)~+~...
~+~(-1)^{n-1}~{{d^{n-1}}\over {dt^{n-1}}}\left({{\partial L}\over 
{\partial z^{(n)}}}\right)
\cr
&~\vdots
\cr
p_{m+n}~&=~~~~~~~~~~~~~~~~~~~~~~~~~~~~~~~~~~~~~~~~~~~~~~~~~~~~~~~~
{{\partial L}\over {\partial z^{(n)}}}
\cr}
\eqno(8)
$$
In particular, for the Lagrangian $L(t,a,{\dot a},{\ddot a}, \Phi)$ we have

$$\eqalign{p_1~&=~{{\partial L}\over {\partial {\dot a}}}~-~
                      {d\over {dt}}\left({{\partial L}\over 
                                {\partial{\ddot a}}}\right)
~=~-{6\over {g_4^2}}{{e^\Phi}{\dot \Phi}}({\dot a}^2~+~K)~\equiv~P_a
\cr
p_2~&=~{{\partial L}\over {\partial {\ddot a}}}
\cr
p_3~&=~{{\partial L}\over {\partial{\dot \Phi}}}~=~a^3{\dot \Phi}~\equiv~P_\Phi
\cr}
\eqno(9)
$$

Given the conjugate momenta to $q_1~=~a$, $q_2~=~{\dot a}$ and $q_3~=~\Phi$
using the Ostrogadski method for higher-derivative actions [31][32][33], 
we find that the Hamiltonian is
$$\eqalign{H~&=~p_1~{\dot q}_1~+~p_2 {\dot q}_2~+~p_3 {\dot q}_3~-~L
\cr
&~=~-6a({\dot a}^2~+~K)~-~6{{e^\Phi}\over {g_4^2}}{\dot \Phi}({\dot a}^2~+~
K){\dot a}
~+~a^3{\dot \Phi}^2~-~{1\over 2}a^3 (D \Phi)^2
\cr
&~=~-g_4^2P_\Phi^{-1}e^{-\Phi}a^4 P_a~+~{1\over {2a^3}}P_\Phi^2
      ~-~\left[{{g_4^2}\over 6}P_\Phi^{-1}e^{-\Phi}P_a^2 a^3 P_a~-~KP_a^2
      \right]^{1\over 2}
\cr}        
\eqno(10)
$$
given the homogeneity of the scalar field.  In a Lorentzian space-time, a 
differential operator is obtained by making
the substitutions 
\hfil\break
$P_a~\to~-i {\partial\over {\partial a}}$ and
$P_\Phi~\to -i {\partial\over {\partial \Phi}}$. 

The Wheeler-DeWitt equation $H\Psi~=~0$ is an pseudo-differential
equation, which can be transformed into a sixth-order partial
differential equation.

$$\eqalign{-{{g_4^2}\over 6} e^\Phi &\left(a^3{{\partial^4\Psi}\over {\partial 
a^3 \partial \Phi}}~+~
6a^2{{\partial^3 \Psi}\over {\partial a^2 \partial\Phi}}\right)~+~
K~e^{2\Phi}\left({{\partial^3 \Psi} \over 
{\partial a^2 \partial \Phi}}~+~{{\partial^4\Psi}\over {\partial a^2
\partial \Phi^2}}\right)
\cr
&~=~a^4g_4^4 \left[4a^3{{\partial \Psi}\over {\partial a}}~+~a^4
{{\partial^2\Psi}\over {\partial a^2}}\right]
~+~ag_4^2 e^\Phi\left({{\partial^4\Psi}\over {\partial a\partial \Phi^3}}
~-~{{\partial^3\Psi}\over {\partial a \partial \Phi^2}}\right)
\cr
~&+~{3\over 2}g_4^2 e^\Phi \left(a {{\partial^2 \Psi}\over {\partial a \partial
\Phi}}~-~{{\partial^3\Psi}\over {\partial \Phi^3}}\right)
~+~{1\over {4a^6}} e^\Phi {\partial\over {\partial \Phi}}
\left(e^\Phi {{\partial^5 \Psi}\over {\partial \Phi^5}}\right)
\cr}
\eqno(11)
$$

Point symmetries of this equation may be checked with the Kersten programme 
[34], which can be used when the coefficients multiplying the derivatives 
are polynomial functions of the independent variables, and equation (10)
can be cast in the following form through the change of variables 
$w~=~e^\Phi$:

$$\eqalign{ -{{g_4^2}\over 6}w^2 &\left(a^3{{\partial^4 \Psi}\over 
a^2{\partial a^3 \partial w}}~+~6a^2{{\partial^3\Psi}\over {\partial a^3}}
\right)
~+~Kw^3\left(2{{\partial^2\Psi}\over {\partial a^2}}~+~3{{\partial^3 \Psi}
\over {\partial a^2\partial \Phi}}~+~{{\partial^4\Psi}\over {\partial a^2
\partial \Phi^2}}\right)
\cr
~&=~a^4 g_4^4 \left[4a^3{{\partial \Psi}\over {\partial a}}~+~a^4
{{\partial \Psi}\over {\partial a^2}}\right]
~+~ag_4^2 w^3\left(2 {{\partial^3\Psi}\over {\partial a\partial w^2}}
~+~w {{\partial^4 \Psi}\over {\partial a \partial w^3}}\right)
\cr
~~~~&~+~{3\over 2} ag_4^2 w^2 {{\partial^2\Psi}\over {\partial a \partial w}}
~-~{3\over 2} g_4^2 w^2 \left({{\partial \Psi}\over {\partial w}}
~+~3w {{\partial^2 \Psi}\over {\partial w^2}}~+~
w^2{{\partial^3\Psi}\over {\partial w^3}}\right) 
\cr
&~+~{1\over {4a^6}}\biggl(w^8 {{\partial^6 \Psi}\over {\partial w^6}}
~+~15 w^7 {{\partial^5 \Psi}\over {\partial w^5}}~+~65 w^6 {{\partial^4 \Psi}
\over {\partial w^4}}~+~91 w^5 {{\partial^3 \Psi}\over {\partial w^3}}
~+~34w^4 {{\partial^2 \Psi}\over {\partial w^2}}
\cr
~&~~~~~~~~~~~~~~~~~~~~~~~~~~~~~~~~~~~~~~~~~~~~~~~~~+~2w^3{{\partial \Psi}
\over {\partial w}}\biggr)
\cr}
\eqno(12)
$$

The Laplace transform with respect to the variable $\Phi$ gives a mixed
difference-
\hfil\break
differential equation in $a$ and the transform parameter $s$.  Denoting
the Laplace transform of $\Psi(\Phi,a)$ by $\psi(s,a)$, it follows that
a second-order recurrence operator in $s$ and a third-order differential
operator in $a$ act on the transform of the wave function.  

$$\eqalign{-{{g_4^2}\over 6}&a^3 
\left[(s-1){{d^3\psi (s-1,a)}\over {da^3}}-
{{d^3~\Psi(0,a)}\over {da^3}}\right]-g_4^2 a^2
\left[(s-1){{d^2\psi(s-1,a)}\over {da^2}}-{{d^2~\Psi(0,a)}\over {da^2}}
\right]
\cr
&~~~+~K~\left[(s-1)(s-2){{d^2\psi(s-2,a)}\over {da^2}}
~-~(s-1){{d^2\Psi(0,a)}\over {da^2}}~-~{{d^2\Psi^\prime(0,a)}\over
{da^2}}\right]
\cr
~&~=~a^4 g_4^4\left[4a^3{{d\psi(s,a)}\over {da}}~+~a^4 {{d^2\psi(s,a)}\over
{da^2}}\right]
\cr
~&~~~~~+~a g_4^2\biggl[(s-1)^2(s-2){{d \psi(s-1,a)}\over {da}}
~-~(s-1)(s-2){{d\Psi(0,a)}\over {da}}
\cr
~&~~~~~~~~~~~~~~~~~~-~(s-2){{d\Psi^\prime(0,a)}\over {da}}
~-~{{d\Psi^{\prime\prime}(0,a)}\over {da}}\biggr]
\cr
&~~~~~+~{3\over 2}g_4^2a \left[(s-1){{d\psi(s-1,a)}\over {da}}
~-~{{d\Psi(0,a)}\over {da}}\right]
\cr
&~~~~~-~{3\over 2}g_4^2~\biggl[(s-1)^3 \psi(s-1,a)~-~(s-1)^2\Psi(0,a)
~-~(s-1)\Psi^\prime(0,a)~-~\Psi^{\prime\prime}(0,a)\biggr]
\cr
&~+~{1\over {4a^6}}\biggl[(s-2)^5(s-1)\psi(s-2,a)~-~(s-2)^4(s-1)\Psi(0,a)
      \cr
&~~~~~~-~(s-2)^3(s-1)\Psi^\prime(0,a)~-~(s-2)^2(s-1)
\Psi^{\prime\prime}(0,a)~-~(s-2)(s-1)\Psi^{\prime\prime\prime}(0,a)
\cr
&~~~~~~~~~~~~~~~~~~~~~~~~~~~~~~~~~~~~~~~~~~-~(s-1) \Psi^{(iv)}(0,a)~-~
             \Psi^{(v)}(0,a)\biggr]			     		
\cr}
\eqno(13)
$$

The mixed difference-differential equation can be solved by considering the
differential and recursion operators separately.  Solving first the
recursion relation gives rise to a higher-order ordinary differential
equation in $a$.  Arbitrary parameters are determined by specifying values
of $\Psi(0,a)$ and $\Psi(\Phi_0,a)$, $\Phi_0\gg 1$ and derivatives
up to fifth order in $\Phi$.  The value $\Phi~=~0$ leads to a
vanishing kinetic term for the dilaton field and the quadratic curvature
term being a topological invariant, so that $\Psi(0,a)$ 
should equal the wave function in the minisuperspace of 
Friedmann-Robertson-Walker metrics in a theory of pure gravity.

The path integral which defines the wave function does not
converge when the action in the weighting factor represents Einstein gravity
coupled to a scalar field.  One may anticipate that a higher-derivative
action can be embedded in a renormalizable theory, and that the evaluation
of the path integral with different initial data, determined by the
Hartle-Hawking or tunneling boundary condition, produces a wave function
which then could be compared with inflationary cosmology. This comparison
can be used to select the most appropriate boundary condition.

Standard inflationary cosmology might receive corrections from two sources.
First, the inclusion of graviton loops will alter the perturbative 
calculations.  Secondly, it is appropriate to specify a boundary between
the Planck era and the inflationary epoch and to study quantum effects
on a manifold with a boundary.  The boundary will affect the range of the
transform variable in the momentum space representation, and extrinsic
curvature terms will be relevant for the quantum theory.

The same techniques can be applied to other superstring effective actions
with higher-order curvature terms or theories derived from the
recently-developed M-theories.  The conjugate momenta, the
Hamiltonian and the Wheeler-DeWitt equation may be derived, 
and the solutions to the partial differential equation for $\Psi$ can be 
obtained. Coupling of the scalar field to the curvature tensor should 
generally lead to a form of the Hamiltonian containing inverse 
or fractional powers of the momenta.  When the coefficients of the 
derivatives in the Wheeler-DeWitt equation are exponential 
functions, the Laplace transform can be used to reduce the differential 
equation to a mixed difference-differential equation.  Thus, the form of 
superstring effective actions containing higher-order terms with couplings 
involving exponential functions is conducive to the reduction of the 
Wheeler-DeWitt equation to an equation with derivatives of 
fewer variables.  If solutions to the Wheeler-DeWitt equation for a 
consistent unified theory of gravity and the elementary 
interactions can be obtained, they may represent wave functions adequately 
describing cosmology both at Planck scales and during the inflationary epoch.  

\vskip 10pt
\centerline{\bf Acknowledgements}
\vskip 5pt
\noindent
I would like to thank P. Zeitsch for running the Kersten program on a
sixth-order differential equation similar to the Wheeler-DeWitt equation
and Dr H. C. Luckock for useful discussions regarding quantization of 
higher-derivative theories. Equations (10) - (13) have been altered from
the form given in the talk because the inclusion of a factor of
$a^3$ in the momentum $P_\Phi$ gives rise to a different expression for
the Hamiltonian and the resulting Wheeler-DeWitt equation. Research on this 
project has been supported by an ARC Small Grant.
\vskip 10pt
\centerline{\bf References}
\item{[1]} F. M\"uller-Hoissen, Ann. der Physik ${\underline{7}}$ 
(1991) 543 - 557
\item{[2]} M. de Roo, H. Suelmann and A. Weidmann, Phys. Lett. 
${\underline{280B}}$ (1992) 39 - 46
\item{[3]} H. Suelmann, Int. J. Mod. Physics D, Vol. 3, No. 1 (1994) 
285 - 288
\item{[4]} I. Antoniadis, J. Rizos and K. Tamvakis, Nucl. Phys. 
${\underline{415B}}$ (1994) 497 - 514
\item{[5]} P. Kanti, J. Rizos and K. Tamvakis, Phys. Rev. ${\underline{D59}}$
(1999) 083512 
\item{[6]} F. M\"uller-Hoissen, Classical Quant. Grav. 
${\underline{3}}$ (1986) 665 - 682
\item{[7]} J. Demaret, Y. De Rop, P. Tombal and A. Moussiaux,
Gen. Rel. Grav. ${\underline{24}}$(1992) 1169 - 1183.
\item{[8]} G. A. Mena Marugin, Phys. Rev. ${\underline{D42}}$ (1990) 
2607 - 2620
\item{[9]} K. Kleidis and D. B. Papadopoulos, J. Math. Phys. 
${\underline{38}}$(6) (1997) 3189 - 3208
\item{[10]} M. B. Mijic, M. S. Morris, W.-M. Suen, Phys. Rev. 
${\underline{D34}}$ (1986) 2934 - 2946
\hfil\break
 M. B. Mijic, M. S. Morris, W.-M. Suen, Phys. Rev. 
${\underline{D39}}$ (1989) 1496 - 1510
\hfil\break
M. S. Morris, Phys. Rev. ${\underline{D39}}$ (1989) 1511 - 1516
\item{[11]} A. Berkin, Phys. Rev. D ${\underline{D44}}$ (1991) 1020 - 1027
\item{[12]} S. W. Hawking and J. C. Luttrell, Nucl. Phys.
${\underline{B247}}$ (1984) 250 - 265
\hfil\break 
S. W. Hawking and J. C. Luttrell, Phys. Lett. {143B} (1984) 83 - 86
\item{[13]} A. Dobado and A. L. Maroto, Phys. Rev. ${\underline{D52}}$
(1995) 1895 - 1901
\item{[14]} F. D. Mazzitelli, Phys. Rev. ${\underline{D45}}$ (1992)
2814 - 2823
\item{[15]} H. J. de Vega and N. Sanchez, `Lectures on String Theory in 
Curved Space-Times', in ${\underline{String~Gravity~and~Physics~at~the~
Planck~Energy~Scale}}$ (Erice, 1995) 11 - 63
\item{[16]} A. L. Maroto and I. L. Shapiro, Phys. Lett. ${\underline{414B}}$
(1997) 34 - 44
\item{[17]} A. Lukas, B. A. Ovrut and D. Waldram, Phys. Lett. 
${\underline{B393}}$ (1997) 65 - 71; `The Cosmology of M-Theory and Type II
Superstrings', hep-th/9802041; based on talks given at the Europhysics
Conference on High Enrgy Physics, Jerusalem, Israel, August 1997 and at the
XXXIIIrd Rescontres de Moriond, Fundamental Paramters in Cosmology,
Les Arcs, Savoie, France, January, 1998
\item{[18]} A. P. Billyard, A. A. Coley, J. E. Lidsey, U. S. Nilsson,
`Dynamics of M-Theory Cosmology' (1999) hep-thy/99080102
\item{[19]} S. W. Hawking, `Quantum Cosmology', ${\underline{Relativity,~
Groups~and~Topology~II}}$, 
\hfil\break
Les Houches 1983, Session XL, edited by
B. S. De Witt and R. Stora (North-Holland, Amsterdam, 1984) 333 - 379  
\hfil\break
S. W. Hawking, Nucl. Phys. ${\underline{B239}}$ (1984) 257 - 276
\item{[20]} A. Vilenkin, Phys. Rev. ${\underline{D32}}$ (1985) 2511 - 2548
\item{[21]} A. D. Linde, `Inflation and Quantum Cosmology',
${\underline{Three~Hundred~Years~of}}$
\hfil\break
${\underline{Gravitation}}$, ed. by S. W. Hawking and
W. Israel (Cambridge: Cambridge University Press, 1989) 604 - 630
\hfil\break
A. D. Linde ${\underline{Inflation~and~Quantum~Cosmology}}$, ed. 
by R. H. Brandenburger 
\hfil\break
(Boston: Academic Press, 1990)
\item{[22]} N. A. Lemos, Phys. Rev. ${\underline{D53}}$ (1996) 4275 - 4279
\item{[23]} J. E. Lidsey, Class. Quantum Grav. ${\underline{11}}$ (1994)
1211 - 1224; Phys. Rev. ${\underline{D49}}$ (1994) 599 - 602    
\item{[24]} D. Clancy, J. E. Lidsey and R. Tavakol, Phys. Rev. ${\underline
{D59}}$ (1999) 063511
\item{[25]} R. Brustein and R. Madden, JHEP 9907 (1999) 006
\item{[26]} M. Gasperini, Int. J. Mod. Phys. ${\underline{A13}}$ )(1998)
4779 - 4786
\item{[27]} R. Brustein, M. Gasperini and G. Veneziano, Phys. Rev. 
${\underline{D55}}$ (1997) 3882 - 3885
\item{[28]} M. D. Pollock, Nucl. Phys. ${\underline{B315}}$ (1989) 
528 - 540;
Nucl. Phys. ${\underline{B324}}$ (1989) 187 - 204
\item{[29]} M. D. Pollock, Int. J. Mod. Phys. A ${\underline{7}}$(17)
(1992) 4149 - 4165; Int. J. Mod. Phys. D ${\underline{4}}$(3) (1995) 
305 - 326
\item{[30]} H. van Elst, J. E. Lidsey and R. Tavako, Class. Quantum Grav.
${\underline{11}}$ (1994) 2483 - 2497
\item{[31]} M. Ostrogradski, Mem. Acad. Imp. Sci. St. Petersburg, Serie VI
(1850) 385
\item{[32]} E. T. Whittaker, ${\underline{A~Treatise~on~the~
Analytic~Dynamics
~of~Particles~Rigid~Bodies}}$ (Cambridge: Cambridge University Press, 1927)
pp. 265 - 267
\item{[33]} M. D. Pollock, Mod. Phys. Lett. ${\underline{A12}}$ (1997)
2057 - 2064
\item{[34]} P. H. M. Kersten (1985) ${\underline{Infinitesimal~Symmetries:
~A~Computational~Approach}}$, Ph. D. Thesis, Technische Hogeschool Dekanen

\end